\documentclass[a4paper,11pt]{article}
\pdfoutput=1 % if your are submitting a pdflatex (i.e. if you have
\usepackage{jinstpub} % for details on the use of the package, please
\usepackage{subcaption}
\usepackage{lineno}
\usepackage{float}
\usepackage{natbib}
\usepackage{enumitem}
\usepackage{upgreek}
\setlist{nolistsep}

\title{\boldmath  Defect detection and size classification in CdTe samples in 3D }

\author[a,b,1]{M. Väänänen,\note{Corresponding author.}}
\author[b]{M. Kalliokoski,}
\author[b]{R. Turpeinen,}
\author[b]{M. Bezak,}
\author[a]{P. Luukka,}
\author[a]{A. Karjalainen,}
\author[a]{A. Karadzhinova-Ferrer}

\affiliation[a]{Lappeenranta-Lahti University of Technology,\\Yliopistonkatu 34, Lappeenranta, Finland}
\affiliation[b]{Helsinki Institute of Physics,\\Gustaf Hällströmin katu 2b, Helsinki}

\emailAdd{mika.vaananen@lut.fi}

\abstract{Defects in semiconductor crystals can have significant detrimental effects on their performance as radiation detectors \cite{cdte_materials}. Defects cause charge trapping and recombination, leading to lower signal amplitudes and poor energy resolution \cite{Golovleva_2021}. We have designed and built a modular 3D scanner for analyzing these defects in semiconductor samples using commercial off-the-shelf components. Previous solutions offer great spatial resolution, but have limited sample holding capacity and use continuum light sources which can cause difficulty differentiating between different materials within samples \cite{alex_master_thesis}. Our design also includes a modular sample holder allowing for easy changing of samples. In this paper, we showcase first results achieved with this custom built scanner as well as planned developments.}

\proceeding{25$^{\text{th}}$ International Workshop on Radiation Imaging Detectors\\
  30 June - 4 July 2024\\
  Lisbon, Portugal}

\begin{document}
\maketitle
\flushbottom

\section{Introduction} \label{sec:intro}
Cadmium telluride (CdTe) is a high Z material which makes it an exceptional choice for gamma sensitive radiation detectors. Its wide band gap of 1.44 eV also makes it suitable for use in room temperature devices without the need for additional cooling \cite{progress_in_development_of_cdte}. However, since the material is more difficult to grow than silicon, the crystals often have comparatively more defects, like tellurium (Te) inclusions \cite{cdte_processing}. These defects in the crystals can negatively affect the performance of the devices in many applications. In radiation detectors, defects manifest as unwanted behaviors: impurities may introduce new energy levels in the band gap causing charge trapping and decreased charge mobility, and larger effects, such as inclusions, grain boundaries and twins can lead to areas of higher or lower conductivity compared to the pure material \cite{correlation_of_Te_inc}. Such effects can deteriorate the count rate, the spectral performance and the lifetime of the device. As such, it's extremely important to monitor the quality of the crystals to make the best use of the material or improve the growth process. Infrared (IR) microscopy is a great tool for such quality control studies, since different materials have different IR transmission spectra. CdTe is partially transparent to IR light from 850 nm and up, while both Cd and Te inclusions are opaque \cite{cdte_inclusions_ir_contrast}. 

Utilising this principle, we have designed and constructed a 3D optical scanner to image and detect defects in semiconductor materials. The base of the scanner is a commercial desktop computer numerical control (CNC) milling machine, modified by replacing the spindle with a custom sample holder and adding a light source and a camera. We show typical images produced by the system before and after image processing. Using OpenCV, we can extract the locations and sizes of defects in samples from the images. We also describe a technique to remove duplicately detected defects from overlapping images and identify the depth of each defect, allowing a full 3D reconstruction of the defect locations within the sample and estimating a size distribution. We have tested it on CdTe samples, and in this paper, we demonstrate the capability to classify the defects based on size and location within the sample.

\section{Scanner design} \label{sec:scanner_design}
\subsection{3D stage}\label{sec:3d_stage}
\begin{figure}[H]
\centering
    \begin{subfigure}{0.35\textwidth}
        \centering
        \includegraphics[width=0.95\linewidth, trim={2cm 5cm 0 1cm}, clip]{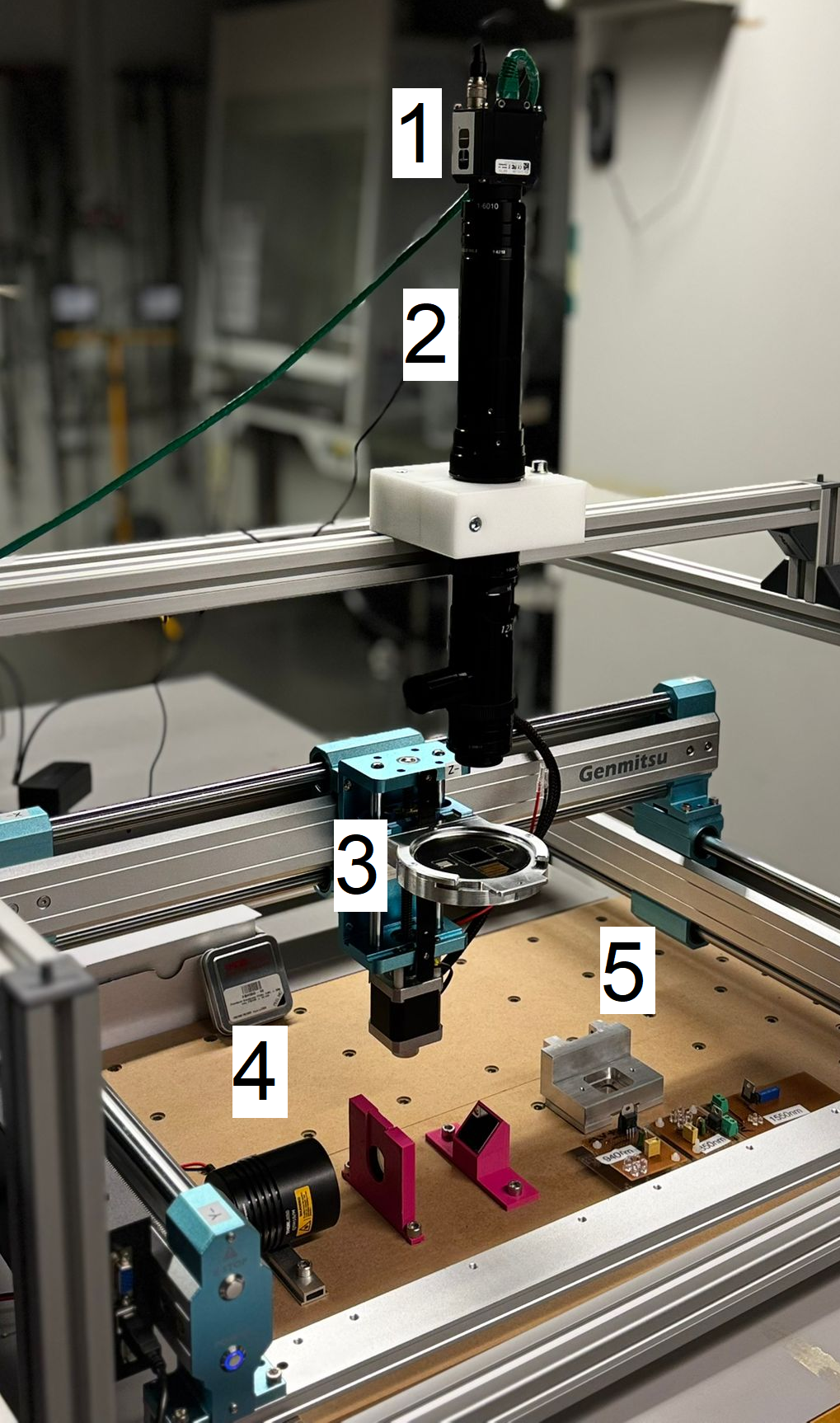}
        \caption{ }
        \label{fig:scanner_overview}
    \end{subfigure}%
    \begin{subfigure}{0.6\textwidth}
        \centering
        \includegraphics[width=\linewidth, trim={0cm 0cm 0cm 0cm}, clip]{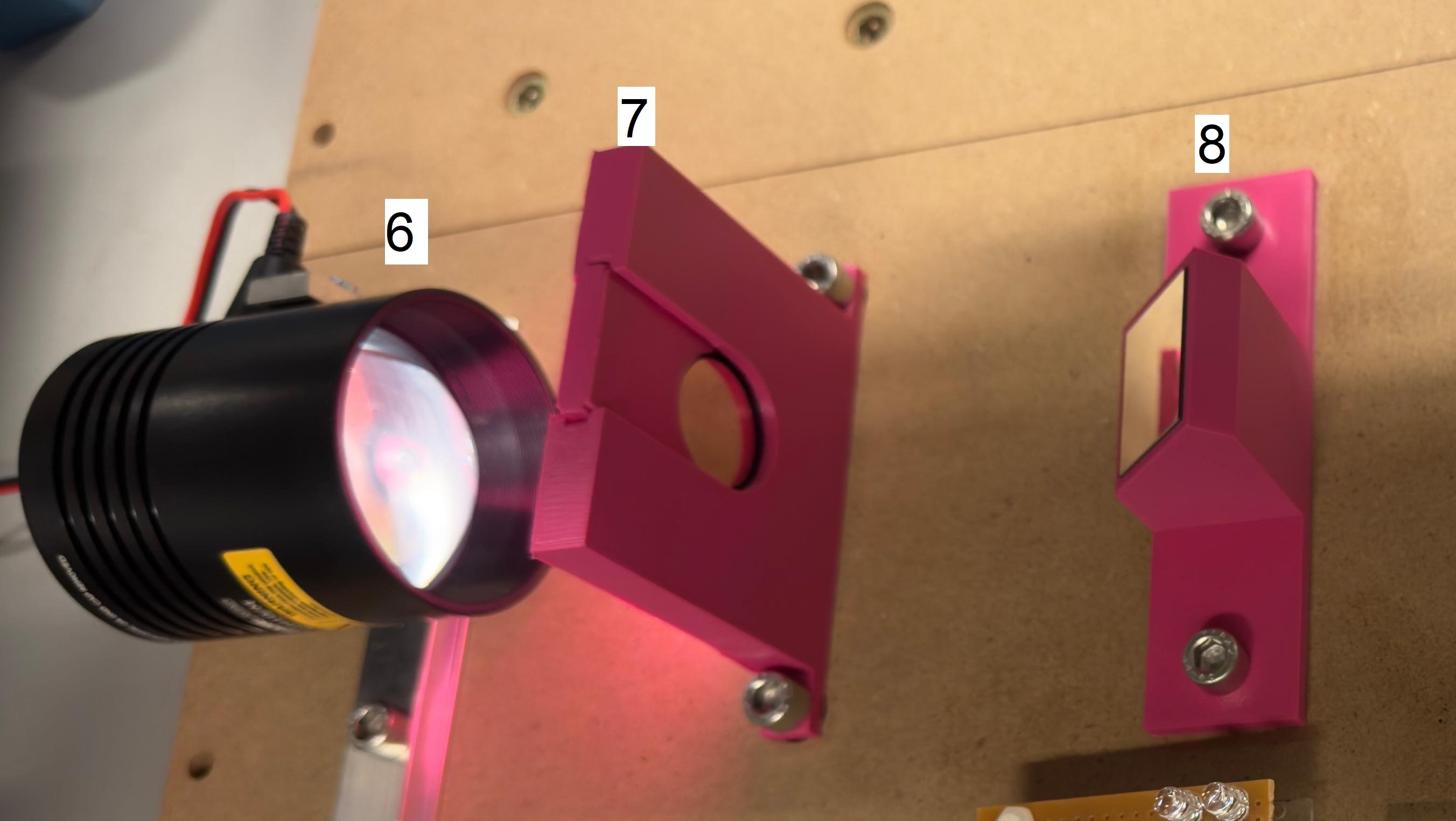}
        \caption{ }
        \label{fig:scanner_light_source}
    \end{subfigure}
    \caption{a) The scanner system and its components: camera (1) objective (2),  wafer holder (3), light source (4) and small 10 mm by 10 mm sample holder (5). b) Closeup of the light source setup: halogen lamp (6), filter holder (7) and mirror holder (8)}
    \label{fig:scanner}
\end{figure}

The scanner was designed around a Genmitsu 4040-PRO desktop CNC milling machine as a 3D sample holding stage \cite{cnc_product_page}. The tool holder and the spindle of the machine were replaced with a modular sample holder, while a camera and a light source are fixed in place above and below the sample holder respectively. The scanner is shown in figure \ref{fig:scanner}. The samples are moved along the three axis, allowing to scan the full volume of the sample. The sample holder (3) is fixed on the Z stage and new sample holders for different samples can easily be manufactured or even 3D printed. So far, the largest sample we have scanned is a 2" wafer, but samples up to 200 mm in diameter can be scanned. By using optics with a shallow depth of field, we can focus the camera to small objects inside the samples, similar to the scans in \cite{stefanie_ir_nn}. The scan proceeds from the top right corner of the sample to the bottom left, row by row and layer by layer. The scanning process is depicted in figure \ref{fig:scan_geometry}.

\subsection{Optics}\label{sec:optics}
One of the novelties of the scanner lies in the adaptability of the light source (4). The wavelength used for the scans can easily be changed, allowing for scans of different materials. The optics of the scanner system consist of an IDS uEye UI-5490SE monochrome camera (1), a Navitar 12x lens system with adjustable magnification (2) and an IR light source. The camera used in this study has a 10 megapixel sensor and is sensitive up to 1000 nm wavelengths, but can easily be changed depending on the application. The magnification of the optics range from 0.58x to 7x, and the depth of field from 50 $\upmu $m to 1.39 mm \cite{navitar_specs}. The light source is a Thorlabs halogen white light source and the wavelengths are changed using band-pass filters \cite{thorlabs_lamp, thorlabs_filters}. A closeup of the light source setup is shown in \ref{fig:scanner_light_source}. The resolving power of the system was tested with a USAF1951 optical testing standard target \cite{usaf1951}. The maximum resolving power according to the test chart is approximately 3 $\upmu$m, which allows us to detect small Te inclusions in CdTe.

\subsection{Control software}\label{sec:control_sw}
Both the camera and the 3D stage are controlled with a software written in Python. The samples are moved in steps with a sufficient time in between for picture taking. Analysis is later performed on a separate computer.

\begin{figure}
    \centering
    \includegraphics[width=0.5\linewidth]{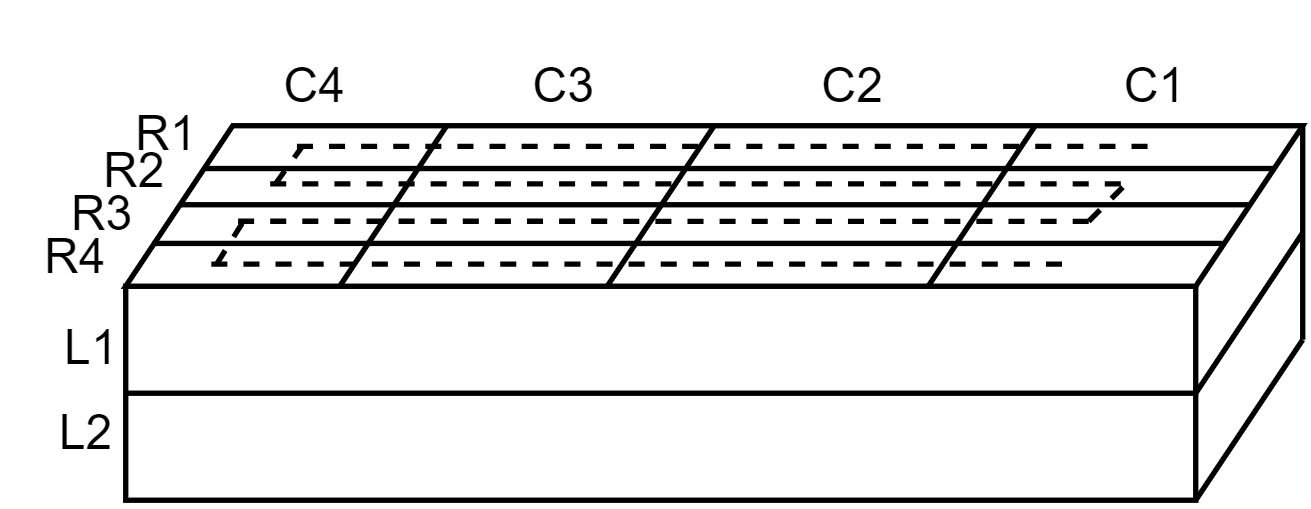}
    \caption{The scan pattern. The volume of the sample is divided into layers, and each layer is divided into a grid. Each grid cell is the size of the image area of the camera, and each layer is the thickness of the optics depth of field. The scan starts from the top right cell (C1, R1, L1 in the diagram) and proceeds row by row, layer by layer until the bottom left corner (C4, R4, L2 in the diagram). The dashed line illustrates the scan pattern. Actual amount of cells and layers depends on the size of the sample and magnification of optics.}
    \label{fig:scan_geometry}
\end{figure}

Before starting a scan, some settings are configured. The control software needs to know the size of the sample, and the number of images to be taken over each direction. Ideally, the sample would be moved by the size of the image in each dimension and afterwards the images could just be tiled. However, in reality, all the components in the scanner are slightly misaligned, which makes the simple tiling approach practically impossible. Thus the aim is to cover the area of the sample so there is at least 50\% overlap between images to enable reliable stitching in later stages of the analysis. This is achieved by moving the sample by half the image width or length, depending on which direction the sample is moved in. For example, if the image area is 1 mm horizontally by 0.8 mm vertically, the sample will be moved by 0.5 mm and 0.4 mm each step respectively. The exact image area depends on the magnification of the optics, which is set with a manual dial on the objective. The scan pattern is shown in figure \ref{fig:scan_geometry}. The scans are started at the top layer, top right corner and progress from the right to left and top to bottom.

After configuring the sample geometry, the exposure time and the pixel clock of the camera are set. After the exposure setting, the scanner is first focused on the top surface of the sample. When the surface is found, the top right corner of the sample is moved to the middle of the image area as a starting point of the scan. The exposure, the sample size and the step size settings are then saved in a log file.

\section{Image processing} \label{sec:processing}
\subsection{Preprocessing}\label{sec:preprocessing}
\begin{figure}
    \centering
    \begin{subfigure}{0.35\linewidth}
        \centering
        \includegraphics[width=\linewidth]{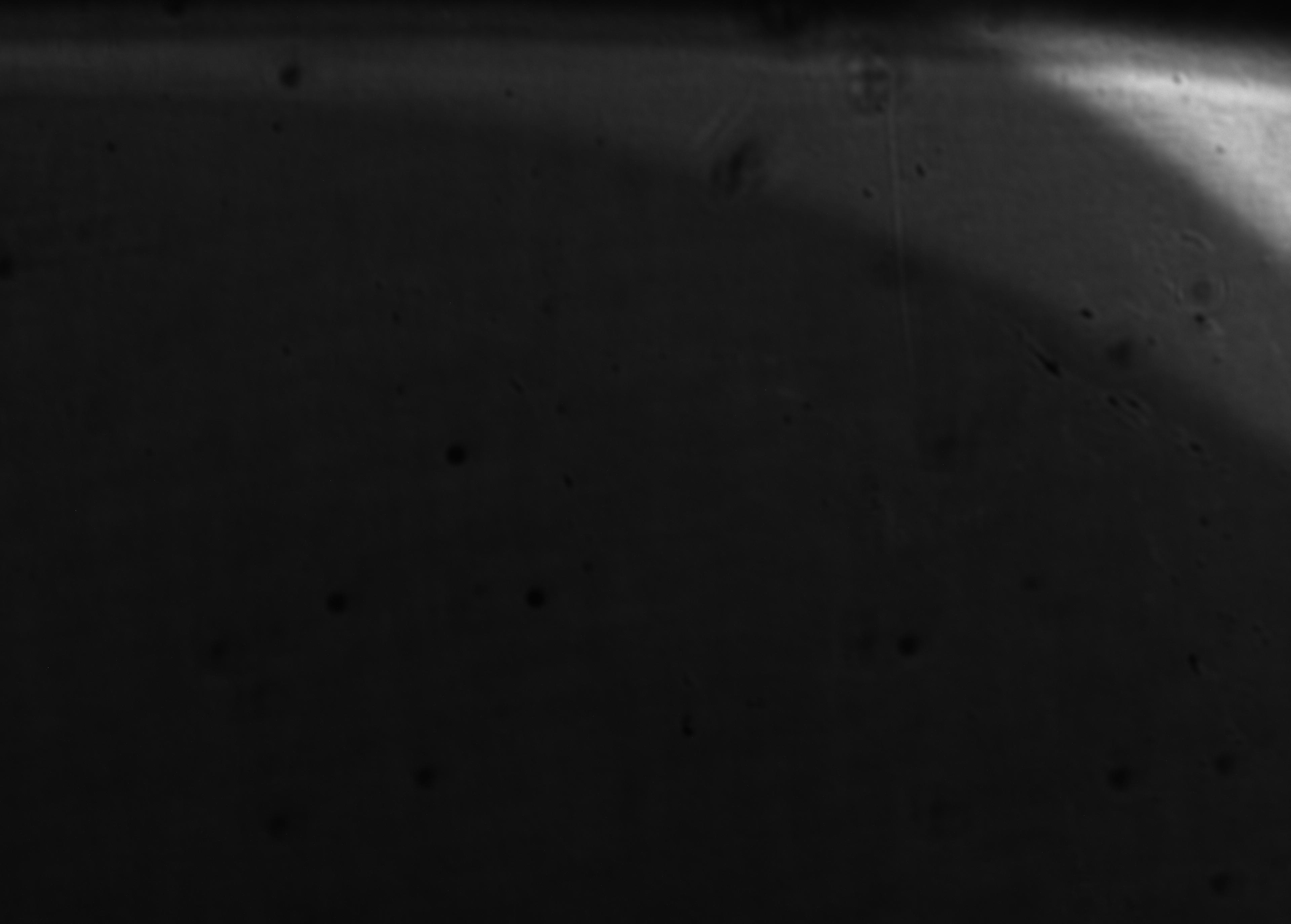}
        \caption{Before processing}
        \label{fig:cdte_gold_no_processing}
    \end{subfigure}%
    \begin{subfigure}{0.35\linewidth}
        \centering
        \includegraphics[width=\linewidth]{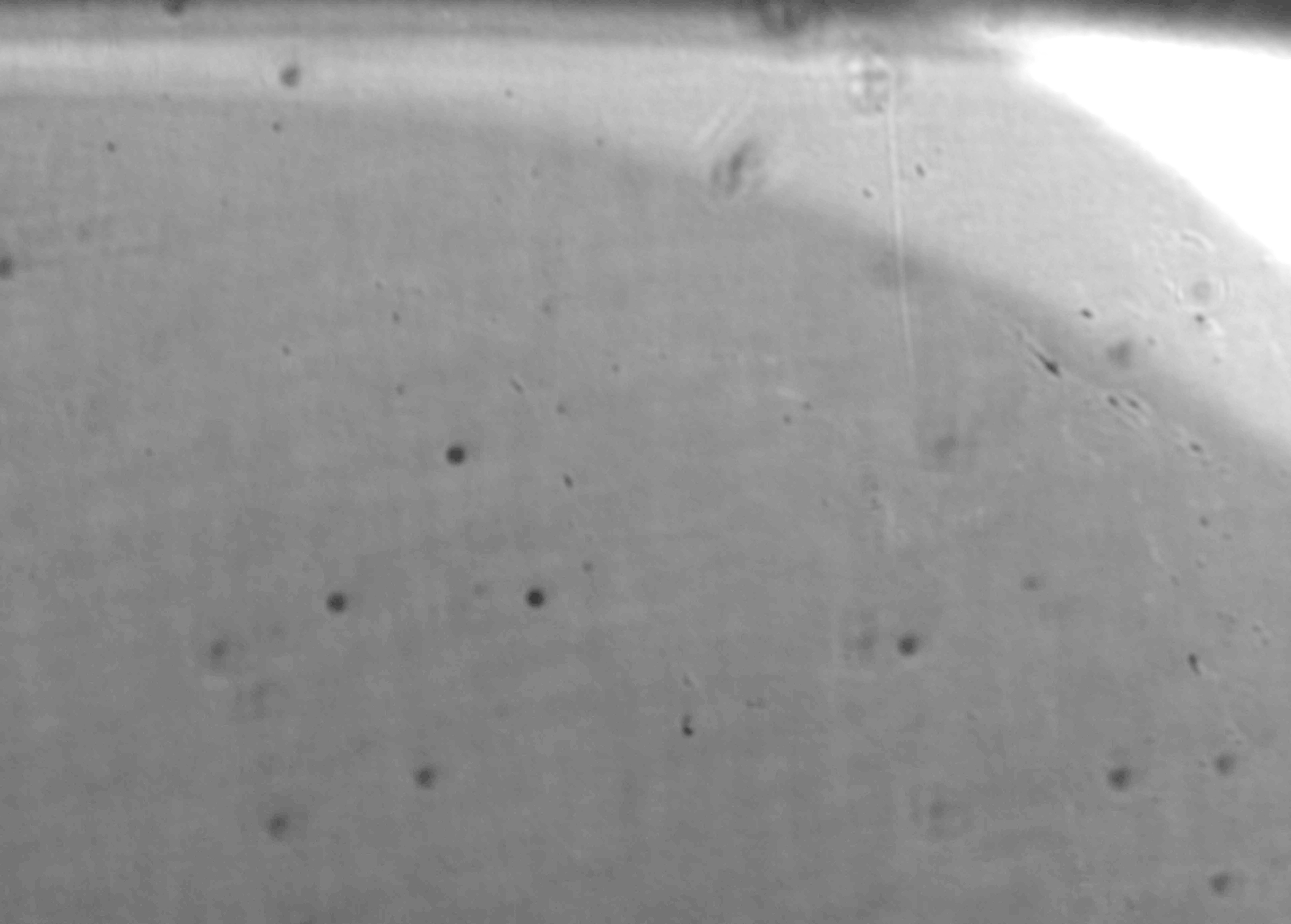}
        \caption{After processing}
        \label{fig:cdte_gold_processing}
    \end{subfigure}
    \caption{Effect of preprocessing on feature visibility in images. The sample shown is a 1 mm thick CdTe crystal, coated on both sides with gold. The image on the left shows a raw image taken with 950 nm light directly from the scanner. Most features are barely visible to the naked eye and difficult to identify with the analysis software. The image on the right shows a processed image, where the brightness and contrast of the image were increased and a gamma correction was applied. This brings out much more features and enables further analysis.}
    \label{fig:gold_cdte_compare}
\end{figure}

After scanning a sample, most images require some processing before further analysis. The preprocessing procedure used in this study consists of a brightness, contrast and gamma correction for exposure correction and Gaussian blur for noise removal. The preprocessing steps required depend on the samples themselves: for example, the sample shown in figure \ref{fig:gold_cdte_compare}  was coated with gold on both sides, reducing light transmission drastically. However after applying the full preprocessing procedure, many features can be detected in the images even with the naked eye.

\subsection{Feature detection}\label{sec:feature_detection}
After preprocessing, further analysis of the images to identify features in the samples. For this we have employed the OpenCV library for Python \cite{opencv_library}. By scanning the processed images one by one, the inclusions in each image were detected using the \texttt{SimpleBlobDetector} class, which is designed to extract blobs in images. The detector binarizes the image, leaving dark inclusions as black dots on a white background, and records the sizes and locations of each dot. The detector also allows filtering of the detected features by size, which helps reducing false positives and speeds up the analysis. After detection, the size and location of each feature is saved for further analysis.

Since the analysis script goes through the images one by one, features in the overlapping sections of images will inevitably be double counted. These double count features can be filtered out by using the sample size and step size between images recorded in the log file by the scanner to calculate the overlap between images, or by stitching the images to create a coordinate map of the defect locations.

\subsection{Depth detection}\label{sec:depth_mapping}
\begin{figure}[H]
    \centering
    \begin{subfigure}{0.35\linewidth}
    \includegraphics[width=\linewidth,trim={0 0 30cm 25cm},clip]{cdte_images/surface_preproc.jpeg}
    \caption{Sample surface}
    \end{subfigure}%
    \begin{subfigure}{0.35\linewidth}
    \includegraphics[width=\linewidth,trim={0 0 30cm 25cm},clip]{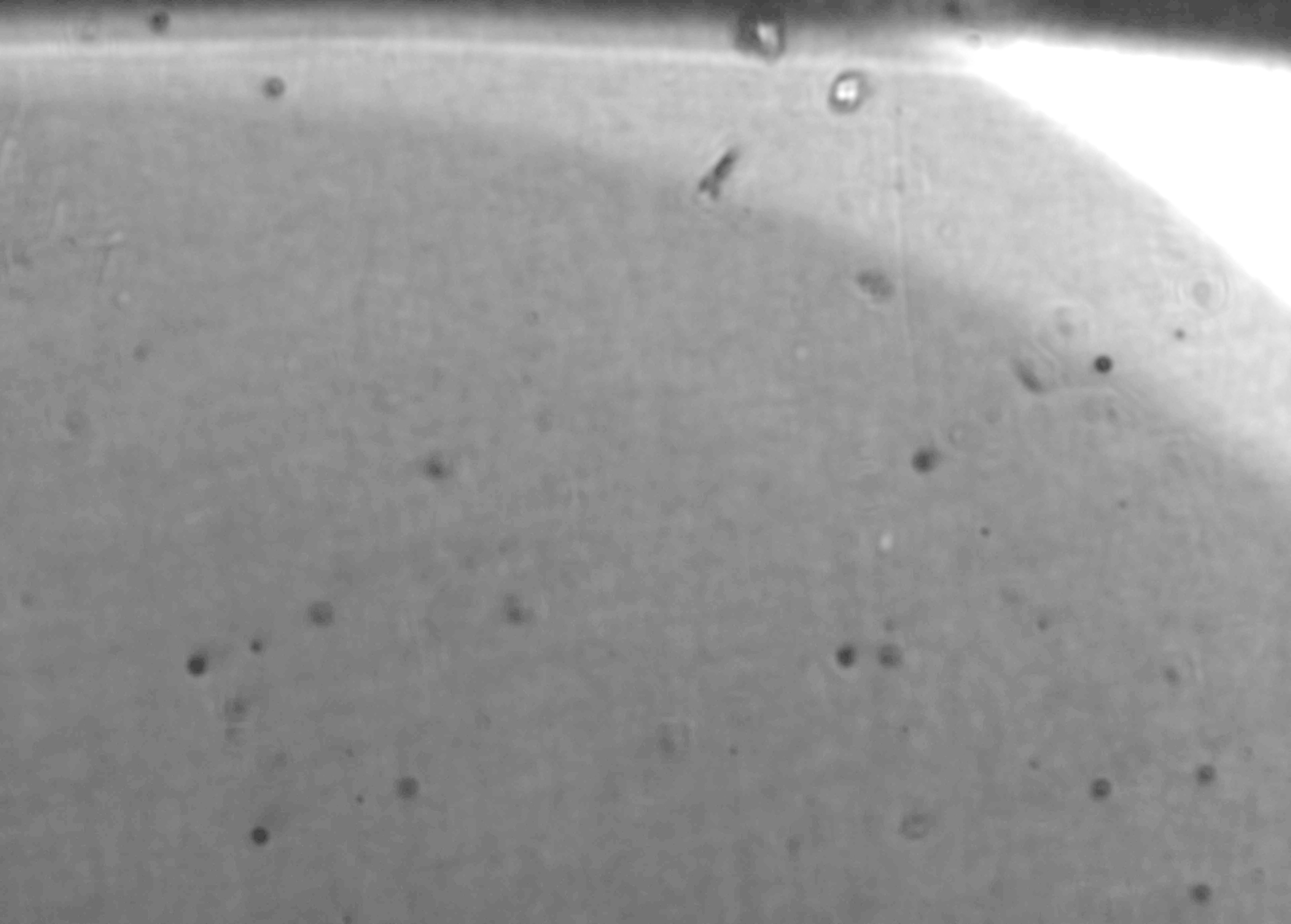}
    \caption{500 $\upmu $m depth}
    \end{subfigure}
    \caption{Comparison of two images of the same area of the sample with the camera focused at different depths. The sample is the same as shown in figure \ref{fig:gold_cdte_compare}, but zoomed in. Most features can be seen in both images, but they are sharp and in focus only in one.}
    \label{fig:cdte_depth_comp}
\end{figure}

\begin{figure}
    \centering
    \begin{subfigure}{0.4\textwidth}
    \includegraphics[width=\linewidth,trim={0cm 0.5cm 2cm 1.3cm},clip]{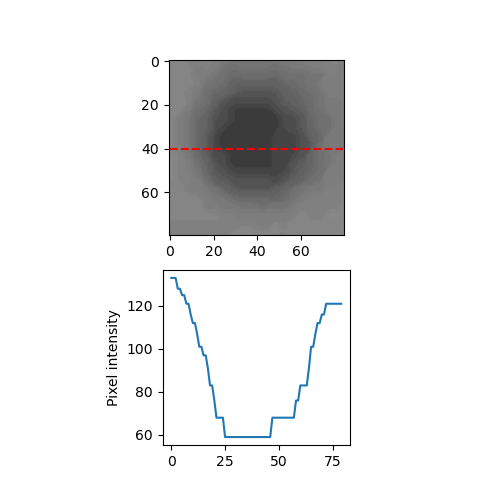}
    \caption{Sample surface in focus}
    \label{fig:top_gradient}
    \end{subfigure}%
    \begin{subfigure}{0.4\textwidth}
    \includegraphics[width=\linewidth,trim={2cm 0.5cm 0cm 1.3cm},clip]{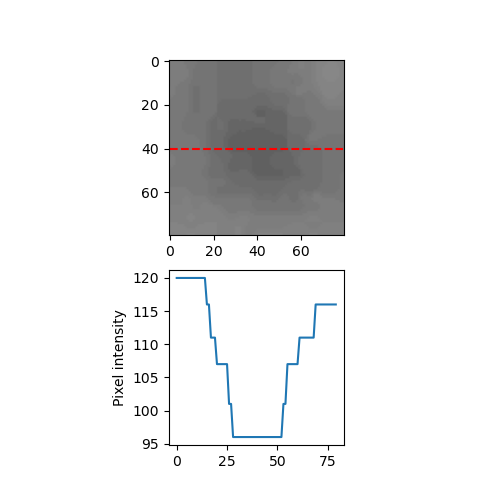}
    \caption{Depth 500um in focus}
    \label{fig:depth_gradient}
    \end{subfigure}
    \caption{Comparison of the intensity of a defect imaged at two depths. a) Defect imaged with focus at surface and the intensity projection along the red dashed line. b) Defect imaged with focus at 500 $\upmu $m depth and the intensity projection along the dashed line. While the defect is visible in both images, the intensity gradient is larger in the surface focus image. Thus, the defect was determined to be on the surface of the sample.}
    \label{fig:defect_gradient_comparison}
\end{figure}

The 3D location of features in the samples can also be reconstructed from the scanner images. While the depth of focus of the optics is narrow, out of focus objects are still visible in the images, as shown in figure \ref{fig:cdte_depth_comp}. This allows us to see objects inside the sample volume even if we have focused the optics on the top surface. After running feature detection on all images, we can stack them at different layers and follow each defect through the stack. The central location of the feature is the depth where it appears the sharpest. The sharpness is calculated from the gradient of the intensity, and the sharpest image is where the gradient reaches the maximum value. The intensity gradients of a defect in and out of focus are shown in figure \ref{fig:defect_gradient_comparison}. In this example, the defect was determined to be at or near the surface of the sample, since the intensity gradient was smaller when the optics were focused at a deeper location.

\section{Results} \label{sec:results}
\subsection{Defect size classification}
\begin{figure}
\centering
    \begin{subfigure}{0.4\textwidth}
    \includegraphics[width=\linewidth,trim={0 0.5cm 0 0}, clip]{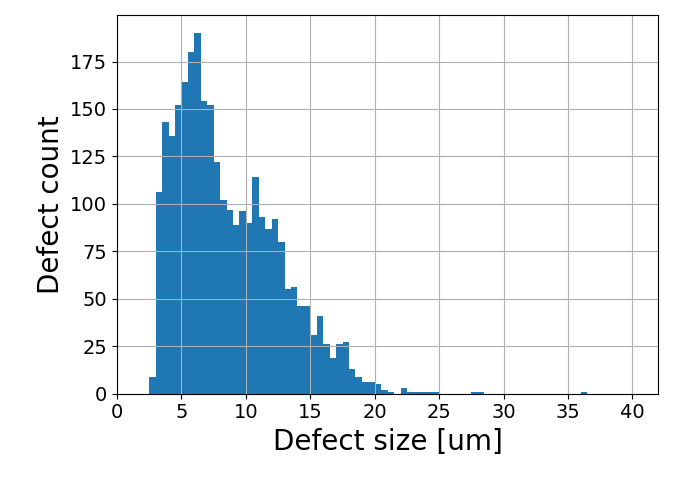}
    \caption{}
    \end{subfigure}%
    \begin{subfigure}{0.45\textwidth}
    \includegraphics[width=\linewidth]{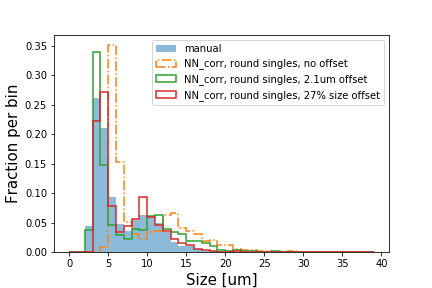}
    \caption{}
    \end{subfigure}
    \caption{a) Size distribution of defects in CdTe obtained with our scanner. b) Size distribution of defects in CdTe samples obtained in a previous study \cite{stefanie_ir_nn} (© IOP Publishing. Reproduced with permission. All rights reserved). Both distributions show the same basic shapes, with peaks at around 6 and 10 $\upmu $m}
    \label{fig:defect_size_dist_comp}
\end{figure}

The size distribution of the features in the samples show a clear resemblance to a previous study \cite{stefanie_ir_nn}. Size distributions obtained in our study and in \cite{stefanie_ir_nn} are shown in figure \ref{fig:defect_size_dist_comp}. Both distributions show at around 6 and 10 $\upmu $m. This is in good agreement with our method, suggesting that the method is viable and produces comparable results to previous studies. With the current optics the maximum depth resolution is 50 $\upmu $m. Since a finer resolution requires more layers per scan, coarser depth scans are much faster to do and for practical reasons a 100 $\upmu $m or even coarser is often used.

\subsection{3D defect mapping}\label{sec:3d_results}
\begin{figure}[H]
    \centering
    \includegraphics[width=0.6\linewidth]{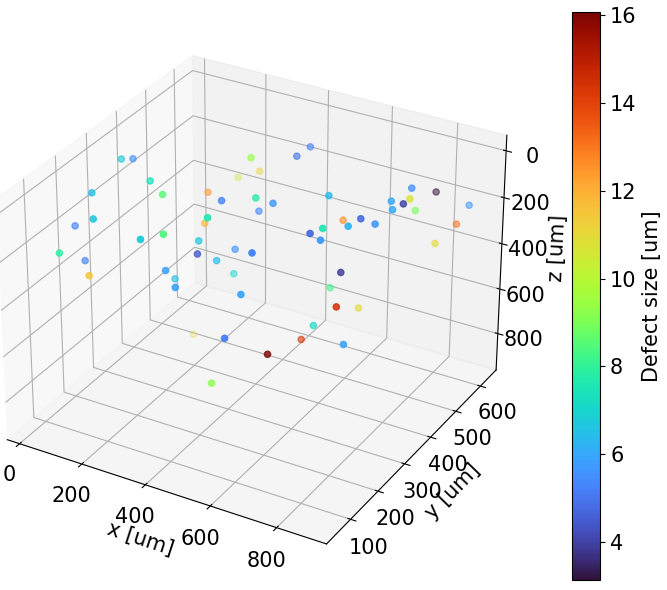}
    \caption{A 3D map of inclusions and their sizes within a section of the sample. Only a small section is shown for clarity.}
    \label{fig:defect_map_3d}
\end{figure}

Based on the depth detection method discussed in \ref{sec:depth_mapping}, we can generate 3D maps of defects in the samples. Figure \ref{fig:defect_map_3d} shows a section of such a map. Only a small section of the whole sample map is shown for clarity. Most of the defects were identified as surface defects. We believe there are a couple of reasons for this: first of all, the scans were done in normal laboratory conditions and not in a clean room, so dust and dirt may have settled on the top surface of the sample. Secondly, various processing techniques, such as chemical treatment on the CdTe crystals can produce more pronounced Te inclusions on the surfaces. Both of these reasons can be checked by repeating the scan in a clean room and by flipping the sample. Stitching of the images would also help in reducing risk of double counting defects.

\section{Conclusions and outlook} \label{sec:outlook}
We have presented the design and the build process of a new transmission scanner for semiconductor characterization. Our scanner allows the user to scan a large size range of samples, up to 3" wafers, and to select the scanning wavelength to tune the scanner on specific materials. We have also demonstrated our analysis software for automatic detection of point-like defects in CdTe samples. We are able to generate 3D maps of defects in the samples and characterize their size distribution. Our software gives us great control over the image processing and feature detection.

In the future, we aim to improve the analysis software significantly. The whole analysis is currently a significant bottleneck in the defect detection pipeline and the code could benefit from eg. parallelization or running on a GPU. Also, larger defects like crystal boundaries, twins and scratches are not currently classified by the software, which is a feature that needs to be implemented. Further work needs to also be done on the stitching of the images to generate large, high resolution images of samples. We also intend to combine the method with other characterization techniques such as transient current technique to study how the defects affect charge generation and transport in different semiconductors.

Furthermore, upgrading the scanner itself would yield even better results. The original 3D stage was used as a proof-of-concept, and as such, is not ideal for this application. Since we are imaging with high-magnification optics, the exposure times are long and any vibrations in the system are very apparent. A more rigid stage with better position repeatability would improve the image quality, scan speeds and stitching process. Additionally, the objective itself could be replaced with an electronically controlled one for repeatable magnification between scans.

\bibliography{iworid_proceedings_bib}

%\appendix
%\section{Appendices}
%Please always give a title also for appendices.

%\acknowledgments
%This is the most common positions for acknowledgments. A macro is
%available to maintain the same layout and spelling of the heading.


\end{document}